\documentclass[prb,twocolumn,showpacs,preprintnumbers,amsmath,amssymb]{revtex4}
\usepackage{graphicx}
\usepackage{dcolumn}
\usepackage{amssymb}
\usepackage{amsmath}

\begin{document}

\title{Charge state of the O$_{2}$ molecule during silicon oxidation through hybrid functional calculations}

\author{Audrius Alkauskas}
\author{Peter Broqvist}
\author{Alfredo Pasquarello}

\affiliation{Ecole Polytechnique F\'ed\'erale de Lausanne (EPFL),
Institute of Theoretical Physics, CH-1015 Lausanne, Switzerland}

\affiliation{Institut Romand de Recherche Num\'{e}rique en Physique
des Mat\'{e}riaux (IRRMA), CH-1015 Lausanne, Switzerland}

\begin{abstract}
We study the charge state of the diffusing O$_2$ molecule during
silicon oxidation through hybrid functional calculations. We
calculate charge transition levels of O$_2$ in bulk SiO$_2$ and use
theoretical band offsets to align these levels with respect to the
Si band edges. To overcome the band-gap problem of semilocal density
fuctionals, we employ hybrid functionals with both predefined and
empirically adjusted mixing coefficients. We find that the charge
transition level $\varepsilon^{0/-}$ in bulk SiO$_2$ occurs at
$\sim$1.1 eV above the silicon conduction band edge, implying that
the O$_2$ molecule diffuses through the oxide in the neutral charge
state. While interfacial effects concur to lower the charge
transition level, our estimates suggest that the neutral charge
state persists until silicon oxidation.
\end{abstract}

\date{\today}


\maketitle

Determining the nature and origin of point defects at
semiconductor-oxide interfaces is a major step in the development of
electronic devices.\cite{Sze} Since direct experimental
characterization of interfaces is often difficult, simulation
methods based on density functional theory (DFT) are increasingly
being used as an alternative tool for the investigation of
atomic-scale properties. However, the study of defect levels
within standard DFT schemes is still largely hindered by the
well-known band-gap problem. Indeed, calculated band gaps and band
offsets are generally severely underestimated and the alignment of
defect levels with respect to the relevant band edges is
consequently affected. A correct alignment is a prerequisite for
understanding issues such as oxide charging,\cite{Robertson_RPP_2006}
 defect passivation, Fermi-level pinning,\cite{Broqvist_APL_2008}
stress-induced leakage current,\cite{Bloechl_PRL_1999} charge state
of defects, etc.

Silicon oxidation is a fundamental process in which charging issues
are expected to play a key role.\cite{Sofield_SST_1995} Our
understanding of this process relies to a large extent on the model
proposed by Deal and Grove, in which oxide growth proceeds through
the diffusion of the oxygen molecule across the oxide followed by
its incorporation at the silicon substrate.\cite{Deal_JAP_1965}
While supporting this general picture, DFT calculations have been
instrumental to achieve an atomic scale description of various
aspects of silicon oxidation, such as the diffusion mechanism,
\cite{Bongiorno_PRL_2002, Bakos_PRL_2002} the oxidation reaction,
\cite{Bongiorno_PRL_2004,Orellana_PRL_2001} the role of spin,
\cite{Kato_PRL_1998} the layer-by-layer oxidation,
\cite{Watanabe_PRL_1998,Tsetseris_PRL_2006} and the release of interstitial
silicon.\cite{Kageshima_PRL_1998} However, the charge state of the
O$_2$ molecule during the oxidation process has so far remained
elusive. This is expected to determine the nature of the oxygen
species occurring at the interface\cite{Stoneham_PRB_2001} and
consequently the oxidation reaction and the atomic structure at the
Si-SiO$_2$
interface.\cite{Sofield_SST_1995,Bongiorno_PRL_2004,Stoneham_PM_1987}

Experimental attempts to determine the charge state of the oxidizing
molecule rest on the study of the oxide growth kinetics in applied
electric fields, but result in conflicting
conclusions.\cite{Jorgensen_JCP_1962} It should further be noted
that the interpretation of such measurements is not trivial because
of the complexity of the underlying atomic
processes.\cite{Bongiorno_PRL_2002} Theoretical work based on
gradient corrected DFT calculations concluded that the oxygen
molecule diffuses through the oxide in a metastable neutral charge
state, assuming the stable negatively charged state only in the
vicinity of the substrate where electron tunneling can
occur.\cite{Stoneham_PRB_2001} 
This leads to a description of silicon oxidation in which neutral 
and negatively charged oxygen species play competing 
roles.\cite{Stoneham_PRB_2001}
This inference directly stems from the position of the charging 
level of the O$_2$ molecule relative to the silicon band edges.
However, the theoretical determination of such an alignment
is subject to the band-gap problem.

In this work, we determine the charge state of the O$_2$ molecule
during silicon oxidation by locating its charge transition
level with respect to the relevant band edges at the Si-SiO$_2$
interface. To overcome the band-gap problem, we use hybrid density 
functionals which give enhanced band gaps compared to semilocal 
functionals.\cite{Muscat_CPL_2001} 
Our results indicate that the neutral state of the oxygen molecule 
is thermodynamically most stable in bulk SiO$_2$ for electron chemical 
potentials lying in the Si band gap. In the vicinity of the 
substrate, the image-charge interactions and the higher oxide density 
concur to lower the charge transition level. Nevertheless, our 
estimates of their effects suggest that the O$_2$ molecules preserve 
their neutral state until the onset of oxidation.

We considered the class of hybrid density functionals based on the
generalized gradient approximation of Perdew, Burke, and Ernzerhof
(PBE),\cite{PBE} which  are obtained by replacing a fraction
$\alpha$ of PBE exchange  with Hartree-Fock exchange. The functional
defined by $\alpha$=$0.25$ is referred to as PBE0 and has received
some support from theoretical considerations.\cite{PBE0} In our
calculations, core-valence interactions were described through
normconserving pseudopotentials and valence wave functions were
expanded in a plane-wave basis set. The basis set is defined by an
energy cutoff of 70 Ry. The pseudopotentials were generated at the
PBE level and used in all calculations. The electronic-structure
calculations corresponding to large supercells were performed with a
Brillouin-zone sampling restricted to the $\Gamma$ point. For bulk
silicon, we determined the band edges using a converged $k$-point
sampling. The integrable divergence of the Hartree-Fock exchange
term was explicitly treated.\cite{Gygi_PRB_1986} Structural
relaxations were carried out at the PBE level. We used the
\textsc{q}uantum-\textsc{espresso}\cite{QE} and
\textsc{cpmd}\cite{CPMD} codes.

For the Si-SiO$_2$ interface, we used a superlattice structure,
in which a crystalline Si layer (9 monolayers) and an amorphous
SiO$_2$ layer ($\sim$17 \AA) are periodically
repeated.\cite{Giustino_PRL_2005} The interface structure was
designed to incorporate a set of atomic-scale features inferred from
experimental data.\cite{Bongiorno_PRL_2003} The simulation cell in
the transverse directions corresponds to a $\sqrt 8 \times \sqrt 8$
Si(100) interface unit. The model contains 131 Si atoms and 86 O
atoms. For bulk amorphous SiO$_2$, we used a disordered 72-atom
model structure obtained previously.\cite{Sarnthein_PRB_1995}


\begin{table}
\caption{Band gaps of Si ($E_{\rm g}^{\rm Si}$) and SiO$_2$ ($E_{\rm
g}^{\text{SiO}_2}$), and valence ($\Delta E_{\text{v}}$) and
conduction ($\Delta E_{\text{c}}$) band offsets at the Si-SiO$_2$
interface, calculated at the PBE level, at the PBE0 level, and
through the use of the mixed scheme (Ref.\
\onlinecite{Alkauskas_2008b}). The experimental values are taken
from Ref.\ \onlinecite{Himpsel_PRB_1988}.}
\begin{ruledtabular}
\begin{tabular}{l c c c c}
&  $E_{\rm g}^{\rm Si}$   & $E_{\rm g}^{\text{SiO}_2}$ & $\Delta
E_{\text{v}}$ & $\Delta E_{\text{c}}$  \\ \hline
PBE  &  0.6 &  5.4 &  2.5  &  2.3  \\
PBE0 &  1.8 &  7.9 &  3.3  &  2.8  \\
Mixed&  1.1 &  8.9 &  4.4  &  3.4  \\
Expt.&  1.1 &  8.9 &  4.4  &  3.4  \\
\end{tabular}
\end{ruledtabular}
\label{tab1}
\end{table}

First, we addressed the band alignment at the interface, as detailed in 
Ref.\ \onlinecite{Alkauskas_2008b}. Band gaps
calculated at the PBE0 level are generally larger than the PBE
values, but the agreement with experiment is not systematically
improved (Table \ref{tab1}). We determined band offsets by aligning
bulk band extrema through a local reference level determined in the
interface calculation.\cite{Alkauskas_2008b} For crystalline Si,
we took the average electrostatic potential as reference, but used
the semicore O 2$s$ levels for amorphous SiO$_2$ to overcome the
difficulties associated with the structural disorder. As seen in
Table \ref{tab1}, the valence and conduction band offsets calculated
at the PBE level underestimate the experimental results by about 2
and 1 eV, respectively. The use of PBE0, reduces these discrepancies
by a factor of 2. To achieve a better description of the band
alignment, we also considered a mixed scheme in which, for each
interface component, we tuned the Hartree-Fock exchange fraction
$\alpha$ to reproduce the experimental band gap.\cite{Alkauskas_2008b} 
This resulted in $\alpha$=0.11 for Si and $\alpha$=0.34 for SiO$_2$. 
The consistency of this scheme stems from the weak dependence of the 
interfacial dipoles on $\alpha$.\cite{Alkauskas_2008b} 
The band offsets obtained in the mixed scheme show excellent 
agreement with experiment (Table \ref{tab1}).

\begin{figure}
\includegraphics[width=8.0cm]{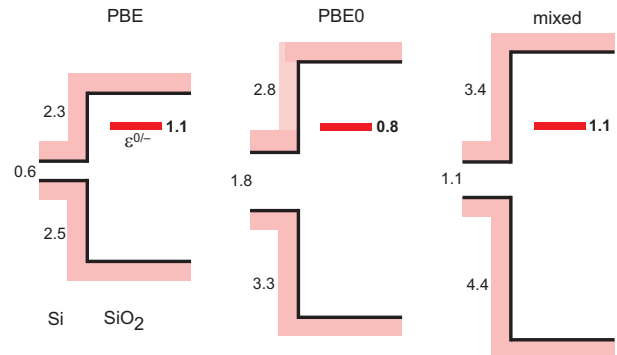}
\caption{(Color online) Charge transition level $\varepsilon^{0/-}$ of the
interstitial O$_2$ molecule in bulk SiO$_2$ aligned with respect to the Si
band edges, as calculated through the PBE, the PBE0, and the mixed scheme.
Values in bold give the position of $\varepsilon^{0/-}$ 
relative to the Si conduction band edge. Other values correspond to the 
Si band gap, the valence band offset, and the conduction band offset. 
Energies are given in eV.}
\label{fig1}
\end{figure}

Second, we determined the relevant charge transition level between the neutral
and the negatively charged state of the interstitial O$_2$ in bulk
SiO$_2$.\cite{Bongiorno_JPCM_2005,Stoneham_PRB_2001} The neutral O$_2$ is
repelled by the oxide network and is found at the center of interstitial
voids.\cite{Bongiorno_PRL_2002} To account for the structural disorder in SiO$_2$,
we considered 10 different interstitial configurations corresponding to
representative local minima of the diffusion path.\cite{Bongiorno_PRL_2002}
The O$_2^-$ attaches to the oxide network by forming a bond to a Si
atom.\cite{Bongiorno_JPCM_2005,Stoneham_PRB_2001} Our study comprises
12 of such atomic configurations. Formations energies of the oxygen molecule
in its charge state $q$ were determined as a function of electron chemical
potential:\cite{VanDeWalle_JAP_2004}
\begin{equation}
E_{\text{f}}^{q}(\mu)=E_{\text{tot}}^{q}-E_{\text{tot}}^{\text{SiO}_{2}}-\
E_{\text{tot}}^{\text{O}_{2}} + q\left(\varepsilon_{\text{v}}+\mu+
\Delta V\right),
\end{equation}
where $E_{\text{tot}}^{q}$, $E_{\text{tot}}^{\text{SiO}_{2}}$, and
$E_{\text{tot}}^{\text{O}_{2}}$ are the total energies of the defect
cell, of the bulk oxide, and of the isolated oxygen molecule, respectively. The electron
chemical potential $\mu$ is referred to the top of the SiO$_{2}$ valence band
$\varepsilon_{\text{v}}$, and $\Delta V$ is the correction needed to align the
electrostatic potential far from the defect to that in the unperturbed bulk
($0.15$ eV in our case). The total energies of the negatively charged species
were corrected for the spurious interaction due to the periodic boundary
conditions.\cite{Makov_PRB_1995} We set the charge transition level $\varepsilon^{0/-}$
in correspondence of the value of the electron chemical potential for which
the average formation energies of the neutral and the negatively charged
oxygen species are equal.\cite{note-sigma} 
Our PBE calculations yielded $\varepsilon^{0/-}$ at 
4.2 eV from the valence band edge, in accord with previous results obtained 
with gradient-corrected functionals.\cite{Bongiorno_PRL_2002, Stoneham_PRB_2001} 
At the PBE0 level, $\varepsilon^{0/-}$ lies at 6.1 eV from the respective 
valence band maximum. In addition, we used a hybrid functional 
with $\alpha$=0.34 in order to reproduce the SiO$_2$ band gap and obtained 
$\varepsilon^{0/-}$ at 6.6 eV from the corresponding valence band maximum.
The empirical adjustment of $\alpha$ is consistent with the mixed scheme 
for the band offsets.\cite{Alkauskas_2008b}. Furthermore,
charge transition levels of the Si dangling bond obtained through such an 
adjustment of the band gap were found to agree with experiment 
within 0.06 eV,\cite{Broqvist_PRB_2008} lending support to this procedure.

Third, we consistently aligned the calculated charge 
transition levels to the Si band edges through the theoretical 
band offsets within each of the three considered schemes.
The resulting alignment is shown in Fig.\ \ref{fig1}. 
The three schemes give a similar picture, 
situating $\varepsilon^{0/-}$ between 0.8 and 1.1 eV above 
the Si conduction band minimum. Since the Fermi level at the
Si-SiO$_2$ interface falls within the Si band gap, 
this result provides convincing evidence that the diffusing 
O$_2$ molecule is thermodynamically most stable in its
neutral charge state.

\begin{figure}
\includegraphics[width=7.5cm]{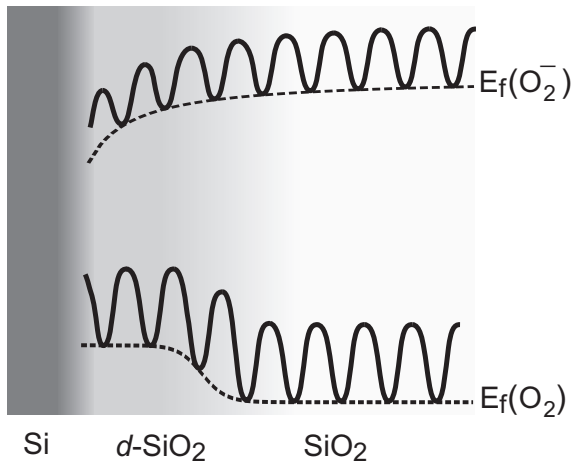}
\caption{Schematic representation of formation energies of the neutral 
(bottom) and negatively charged (top) O$_2$ molecule near the Si-SiO$_2$ 
interface as function of distance to the Si substrate. $d$-SiO$_2$ and
SiO$_2$ represent the densified and the normal oxide, respectively.
The wavy lines pictorially represent the potential energy along the
diffusion path. 
}
\label{fig2}
\end{figure}

We emphasize that this conclusion is naturally drawn provided the
band offsets and the charge transition levels are calculated
within the {\it same} theoretical scheme. For instance, the
use of $\varepsilon^{0/-}$ calculated at the PBE level in combination
with experimental offsets gives a charge transition level falling
below the valence band maximum, leading to a qualitatively different 
description.\cite{Stoneham_PRB_2001} Such an alignment scheme
assumes that the position of the charge transition level with
respect to valence band maximum is already well described in the
semilocal DFT calculation. However, this is generally not the case. 
An improved description of the band gap results in significant 
shifts of both the valence and the conduction bands when referred 
to a reference potential external to the electron 
system.\cite{Shaltaf_PRL_2008,Alkauskas_2008a} 
At variance, charge transition levels of localized impurities 
remain largely unaffected.\cite{Alkauskas_2008a}
This property suggests yet another alignment scheme not relying on
the use of hybrid functionals. Assuming that the charge transition
level calculated at the PBE level does not shift, we first determined
valence band corrections relying on recent $GW$ ($-1.8$ eV) and
quasiparticle selfconsistent $GW$ ($-$2.8 eV)
calculations.\cite{Shaltaf_PRL_2008} The respective $\varepsilon^{0/-}$ 
levels are found at 0.5 eV and 1.5 eV above the silicon conduction band 
edge when aligned through experimental band offsets.
This provides further support for the results in Fig.\ \ref{fig1}.

When the diffusing oxygen molecule approaches the Si substrate, 
two different interfacial effects acting on the neutral and the 
negatively charged O$_2$ molecule need to be considered.
$X$-ray reflectivity measurements indicate the occurrence of a 
densified interfacial oxide, showing a density of 2.4 g/cm$^3$ and 
extending over a distance of about 10 \AA.\cite{Kosowsky_APL_1997} 
The occurrence of a densified oxide affects the incorporation energy 
of the neutral O$_2$ molecule which depends on the size of 
interstitial voids.\cite{Bongiorno_PRL_2002}
Consequently, the average formation energy of neutral O$_2$ in the 
densified oxide increases by 0.6 eV.\cite{Bongiorno_JPCM_2003,Bongiorno_PRL_2002} 
This picture is confirmed by both selected calculations on our interface model 
and recent results in the literature.\cite{Ohta_PRB_2008,Tsetseris_PRL_2006}
Figure \ref{fig2} schematically illustrates this effect showing the evolution 
of the formation energy of the neutral O$_2$ as it approaches 
the interface. In the following, we express the correction with 
respect to the bulk formation energy as $\Delta E^0(z)$, where $z$ 
is the distance to the substrate.

The formation energy of the negatively charged oxygen molecule mainly results
from the formation of a bond with a Si atom of the oxide network and is 
therefore not very sensitive to the modified structural properties of the
near-interface oxide. However, as the negatively charged species approaches 
the silicon substrate, it undergoes stabilization because of the image-charge 
interaction arising from the dielectric discontinuity at the interface. 
From classical electrostatics, the stabilization energy reads
\begin{equation}
\Delta E^-(z) = - \frac{1}{4 z \varepsilon_{\text{SiO}_2}}
\left(\frac{\varepsilon_\text{Si}- \varepsilon_{\text{SiO}_2}}
{\varepsilon_\text{Si}+ \varepsilon_{\text{SiO}_2}} \right),
\end{equation}
where $\varepsilon_\text{Si}$ and $\varepsilon_{\text{SiO}_2}$ are the
static dielectric constants of Si and SiO$_2$, respectively.
This effect is also illustrated in Fig.~\ref{fig2}.

\begin{figure}
\includegraphics[width=7.5cm]{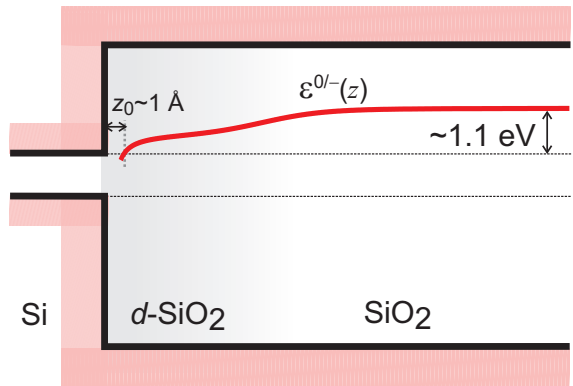}
\caption{(Color online) Evolution of the charge transition level 
$\varepsilon^{0/-}$ for the interstitial O$_2$ molecule as a function of 
distance $z$ to Si substrate. The charge transition level drops below 
the Si conduction band minimum at a distance $z_0 \approx 1$ \AA.
We used $\varepsilon_{\rm{Si}}=$12 and $\varepsilon_{\rm{SiO}_2}$=4.
The densified oxide ($d$-SiO$_2$, shaded) extends over $\sim$ 10 \AA, 
beyond which the oxide (SiO$_2$) recovers its normal density. 
}
\label{fig3}
\end{figure}

As a consequence of the two effects discussed above, the charge transition
level also evolves as the O$_2$ molecule approaches the interface:
\begin{equation}
\varepsilon^{0/-}(z)=\varepsilon^{0/-}_\text{bulk} - \Delta E^0(z) + \Delta E^-(z),
\end{equation}
where $\varepsilon^{0/-}_\text{bulk}$ corresponds to the charge transition level
in bulk SiO$_2$. We remark that both effects concur to lower $\varepsilon^{0/-}$.
To provide a quantitative estimate, we adopted for $\varepsilon^{0/-}_\text{bulk}$ 
the position determined within the mixed scheme, viz.\ at 1.1 eV above 
the silicon conduction band minimum. The resulting evolution of $\varepsilon^{0/-}(z)$ 
is shown in Fig.\ \ref{fig3}. We found that  
the charge transition level lies above the Si conduction band edge as long 
as the distance of the O$_2$ to the Si substrate exceeds $\sim$1 \AA.\cite{note}
At this small distance the oxidation reaction is already under way,\cite{Bongiorno_PRL_2004}
suggesting that the negatively charged oxygen molecule plays only a minor 
role during silicon oxidation. The presence of charged oxygen species had 
previously been invoked to explain the growth kinetics in the thin-oxide 
regime.\cite{Stoneham_PRB_2001} However, alternative mechanisms based on 
a spatially varying diffusion rate do not require the charging of the 
diffusing O$_2$ molecule.\cite{Bongiorno_PRL_2004,Watanabe_PRL_2006} 
In the present picture, the neutral O$_2$ 
interstitial is the dominating oxidizing agent. Other oxygen species 
can only occur upon the reaction of O$_2$ with Si-Si bonds.\cite{Bongiorno_PRL_2004}

In conclusion, we addressed the charge state of the O$_{2}$ molecule during silicon 
oxidation overcoming the band-gap problem through a scheme based on hybrid density 
functionals. The oxygen molecule is found to diffuse in its neutral charge state 
until the onset of oxidation.
More generally, the defect level alignment scheme proposed in this work is expected 
to be very useful for addressing charging issues associated to defects at interfaces.
 
Support from the Swiss National Science Foundation (Grants Nos.\
200020-111747 and 200020-119733) is acknowledged. The calculations
were performed on the Blue Gene of EPFL, at DIT-EPFL and CSCS.


\begin{thebibliography}{10}
\bibitem{note}
Furthermore, the derived distance of $\sim$1 \AA\ is likely 
overestimated because of the divergent nature of the 
image-charge potential in classical electrostatics.
\end{thebibliography}
\end{document}